\documentclass[review,3p,12pt,preprint]{elsarticle}
\usepackage{graphicx}
\usepackage{color}
\usepackage{setspace}
\usepackage{amsmath}
\usepackage{lineno}
\usepackage{epsfig}
\usepackage{alphalph}
\usepackage{float}

\begin{document}


\begin{frontmatter}
\title{Proton Polarimeter Calibration between 82 and 217 MeV}

\author[SMU,Dal]{J.~Glister\corref{label1}}
\cortext[label1]{Corresponding author.}
\ead{jglister@jlab.org}
\author[TelAviv]{G.~Ron}
\author[Seoul]{B.~Lee}
\author[NRCN]{A.~Beck}
\author[Christopher]{E.~Brash}
\author[JLab]{A.~Camsonne}
\author[Seoul]{S.~Choi}
\author[Rutgers]{J.~Dumas}
\author[JLab]{R.~Feuerbach}
\author[JLab,Rutgers]{R.~Gilman}
\author[JLab]{D.W.~Higinbotham}
\author[Rutgers]{X.~Jiang}
\author[JLab]{M.K.~Jones}
\author[NRCN]{S.~May-Tal~Beck}
\author[SMU]{E.~McCullough\fnref{label2}}
\fntext[label2]{Present address: University of Western Ontario, London, Ontario N6A 3K7, Canada}
\author[SouthCarolina]{M.~Paolone}
\author[TelAviv]{E.~Piasetzky}
\author[Ohio]{J.~Roche}
\author[Rutgers]{Y.~Rousseau}
\author[SMU]{A.J.~Sarty}
\author[Virginia,Temple]{B.~Sawatzky}
\author[SouthCarolina]{S.~Strauch}

\address[SMU]{Saint Mary's University, Halifax, Nova Scotia B3H 3C3, Canada}
\address[Dal]{Dalhousie University, Halifax, Nova Scotia B3H 3J5, Canada}
\address[TelAviv]{Tel Aviv University, Tel Aviv 69978, Israel}
\address[Seoul]{Seoul National University, Seoul 151-747, Korea}
\address[NRCN]{NRCN, P.O. Box 9001, Beer-Sheva 84190, Israel}
\address[Christopher]{Christopher Newport University, Newport News, Virginia 23606, USA}
\address[JLab]{Thomas Jefferson National Accelerator Facility, Newport News, Virginia 23606, USA}
\address[Rutgers]{Rutgers, The State University of New Jersey, Piscataway, New Jersey 08855, USA}
\address[SouthCarolina]{University of South Carolina, Columbia, South Carolina 29208, USA}
\address[Ohio]{Ohio University, Athens, Ohio 45701, USA}
\address[Virginia]{University of Virginia, Charlottesville, Virginia 22094, USA}
\address[Temple]{Temple University, Philadelphia, Pennsylvania 19122, USA}

\begin{abstract}

The proton analyzing power in carbon has been measured for energies of 82 to 217 MeV and proton scattering angles of 5 to 41$^{\circ}$.  The measurements were carried out using polarized protons from the elastic scattering $^{1}$H$ ( \vec{e}, \vec{p} )$ reaction and the Focal Plane Polarimeter (FPP) in Hall A of Jefferson Lab.  A new parameterization of the FPP p-C analyzing power was fit to the data, which is in good agreement with previous parameterizations and provides an extension to lower energies and larger angles.  The main conclusions are that all polarimeters to date give consistent measurements of the carbon analyzing power, independently of the details of their construction and that measuring on a larger angular range significantly improves the polarimeter figure of merit at low energies.

\end{abstract}


\end{frontmatter}

\section{Introduction}

In order to extract polarizations from asymmetries in proton-carbon scattering, accurate values of the analyzing power, $A_{c}$, are needed.  The analyzing power depends on the strength of the spin-orbit coupling and is a function of the proton kinetic energy, $T_{p}$, and of the polar scattering angle, $\theta_{fpp}$.  We have used a functional form similar to those used previously~\cite{Ransome1982,McNaughton1985}, to obtain a new parameterization in the energy range of 82 to 217 MeV and proton scattering angles in carbon between 5 and 41$^{\circ}$.  The new parameterization is in good agreement with previous parameterizations~\cite{McNaughton1985,Aprile,Waters,Pospischil} and provides an extension to lower energies and larger scattering angles.  To date, the new parameterization has been used in three low energy polarization experiments~\cite{Ron2007,RonProp,Glister2008}.

\section{Method}

The experiment was carried out in Hall A of Jefferson Lab~\cite{JLab}.  The primary elastic scattering $^{1}$H$ ( \vec{e}, \vec{p} )$ reaction provided the source of polarized protons for the $A_c$ calibration.  The scattered proton polarization was determined independently of our knowledge of $A_c$. 

A continuous, polarized electron beam with polarization ranging from 80 to 85\% was produced using a strained gallium-arsenide (GaAs) source~\cite{Sinclair2007,Hernandez2008}.  The polarization in Hall A was limited to $h$ = 38--41\% due to multi-hall running.  The beam helicity was flipped pseudo-randomly at 30 Hz, with negligible beam charge asymmetry between the two helicity states.  The electron beam was accelerated to either 362 or 687 MeV and then scattered from a 15 cm long liquid hydrogen target.  

The scattered protons were detected in the left High Resolution Spectrometer (HRS)~\cite{Alcorn2004}, made up of one dipole and three quadrupole magnets.  Vertical drift chambers were used to track the protons after deviation by the magnetic field of the dipole, allowing a determination of the momentum with a resolution of 2 $\times$ 10$^{-4}$~\cite{Fissum2001}. The proton target quantities $y_{tg}$ (transverse position), $\theta_{tg}$ ($dx/dz$, where $x$ is the displacement in dispersive plane and $z$ is along the spectrometer axis), $\phi_{tg}$ ($dy/dz$) and $\delta_{tg}$ (fractional deviation of momentum from central trajectory) were reconstructed from the proton trajectory using the HRS optics matrix. Triggering was performed using a coincidence between two scintillator planes, S1 and S2, each made up of six panels.  The scintillators S1 and S2 are both located before the Focal Plane Polarimeter (FPP) carbon analyzer.   

The FPP, downstream of the VDCs and trigger panels, measured the recoil polarization of the protons through a secondary scattering of the protons from a carbon block ``analyzer''.  Spin-orbit coupling between the proton spin and angular momentum about the analyzer carbon nucleus leads to an asymmetry in the azimuthal scattering angle, $\phi_{fpp}$, reconstructed from the front and rear proton tracks. A detailed description of the polarimeter can be found in~\cite{punjabi2005,fpp}.  The FPP measures the inclusive C($p,p^{\prime}$) reaction at the low energies reported here, but also the inclusive C($p,X^{\pm}$), where $X$ is a charged particle, for energies above pion production threshold.

Cuts were made on the reconstructed proton target quantities of $y_{tg} <$ 65 mm, $\theta_{tg} <$ 65 mrad, $\phi_{tg} <$ 38 mrad and $\delta_{tg} <$ 0.045, in order to ensure that the protons originated in the target volume and traversed the HRS in a phase space region with high detection efficiency.  To remove the majority of multiple scattering events due to Coulomb interactions between the proton and carbon nuclei, a cut on the FPP polar angle of $\theta_{fpp} >$ 5$^{\circ}$ was made.  A conetest cut was used to remove instrumental asymmetries~\cite{roche2003}.  In order to ensure that all FPP scattering events originated from within the carbon block, a cut was made around the location (along the spectrometer axis) of closest approach between the front and rear FPP tracks.  A cut of 1 cm or less was also placed on the distance of closest approach between the front and rear tracks.  

Three carbon block thicknesses were used: 0.75, 2.25 and 3.75''.  The carbon thickness was varied according to GEANT~\cite{geant} Monte Carlo studies in order to maximize the figure of merit ($FOM$), which is a measure of how many events contribute to the scattering asymmetry and is given by the following integral over $\theta_{fpp}$ or approximate summation over $N$ bins in $\theta_{fpp}$:

\begin{align}
FOM =  \int_{\theta_{min}}^{\theta_{max}} \epsilon_{fpp}(\theta_{fpp})A^{2}_{c}(\theta_{fpp})d\theta_{fpp} 
    \approx  \sum^{N}_{i=1}{A^{2}_{c,i}\epsilon_{fpp,i}}
\label{eq:fom}
\end{align}
where $A_{c}$ is the analyzing power and $\epsilon_{fpp}$ is the FPP efficiency, given by the fraction of events passing the target cuts and having tracks in the front FPP chambers which pass the FPP cuts.  For the purposes of choosing carbon thicknesses, the Monte Carlo studies assumed $A_{c}$ to follow the earlier parameterization of~\cite{McNaughton1985}.  A summary of the primary $^{1}$H$(\vec{e},\vec{p})$ reaction kinematics and FPP parameters can be found in Table~\ref{tab:kinematics}.  

\section{Analysis}

The recoil transferred polarization was determined by a maximum likelihood method using the difference of the azimuthal distributions at the focal plane corresponding to the two beam helicity states.  After spin transport through the spectrometer using COSY~\cite{cosy}, a differential algebra based code, the transferred polarization products at the target, $hA_{c}P_{x}^{'}$ and $hA_{c}P_{z}^{'}$, were obtained.  The ratio of these products was used to calculate the proton elastic form factor ratio, $G_{E_{p}}/G_{M_{p}}$, with the following equation~\cite{punjabi2005}:

\begin{align}
\frac{G_{E_{p}}}{G_{M_{p}}} = -\frac{hA_{c}P_{x}^{'}}{hA_{c}P_{z}^{'}}\frac{E_{o}+E_{e}}{2M_{p}}\tan\frac{\theta_{e}}{2}
\label{eq:ff_ratio}
\end{align}
where $E_{o}$ and $E_{e}$ are the energies of the incident and scattered electron, respectively, $\theta_{e}$ is the electron scattering angle in the laboratory frame and $M_{p}$ is the proton mass.  Note that knowledge of the analyzing power and beam polarization are not required as they cancel.  The form factor ratio calculated with Equation~\ref{eq:ff_ratio} was in turn used in the following derivations for the transferred polarization components at the target~\cite{theory1,theory2,theory3,theory4}:

\begin{align}
P_{x}^{'} = -2\sqrt{\tau(1+\tau)}\frac{G_{E_{p}}/G_{M_{p}}\tan\frac{\theta_{e}}{2}}{(G_{E_{p}}^{2}/G_{M_{p}}^{2}+\tau/\epsilon)}
\label{eq:pols1}
\end{align}

\begin{align}
P_{z}^{'}=\frac{E_{o}+E_{e}}{M_{p}}\sqrt{\tau(1+\tau)}\frac{\tan^{2}\frac{\theta_{e}}{2}}{(G_{E_{p}}^{2}/G_{M_{p}}^{2}+\tau/\epsilon)}
\label{eq:pols2}
\end{align}
where $\tau = Q^{2}/4M_{p}^{2}$, $Q^{2}$ is the four-momentum transfer squared and $\epsilon = [1+2(1+\tau)\tan^{2}(\frac{\theta_{e}}{2})]^{-1}$ is the longitudinal polarization of the virtual photon.  $P_{x}^{'}$ and $P_{z}^{'}$ calculated with equations~\ref{eq:pols1} and~\ref{eq:pols2}, respectively, were combined with the transferred polarization products at the target obtained through azimuthal asymmetries, $hA_{c}P_{x}^{'}$ and $hA_{c}P_{z}^{'}$, to extract the analyzing power and the statistical uncertainty~\cite{punjabi2005}, where $h$ was taken to be the average beam helicity associated with the data.  

The extracted analyzing power data can be found in Tables~\ref{tab:Ay_362} and~\ref{tab:Ay_687} for incident electron beam energies of 362 and 687 MeV, respectively.  The total systematic uncertainty due to uncertainties in beam polarization, spin transport, reconstruction of FPP scattering angles, proton momentum and beam energy is also 
shown.  Table~\ref{tab:Ay_average} presents the weighted average analyzing power, FPP efficiency and figure of merit, calculated using the summation approximation of equation~\ref{eq:fom} and the data in Tables~\ref{tab:Ay_362} and~\ref{tab:Ay_687}.  

The functional form used to fit the extracted analyzing power is similar to the ``low energy fit'' used by McNaughton et al.~\cite{Ransome1982,McNaughton1985}.  It is given by:

\begin{align}
A_{c}(\theta_{fpp},p_p) = \frac{ar}{1+br^{2}+cr^{4}+dr^{6}}
\end{align}
where $r=p_{p}\sin(\theta_{fpp})$ and $p_{p}$ is the proton momentum in GeV/c at the center of the carbon analyzer.  Note that the functional form given satisfies the physical constraint that the analyzing power vanish at $\theta_{fpp}$ = 0 and 180$^{\circ}$.  The coefficients $a$, $b$, $c$ and $d$ are polynomials of the momentum.  The last parameter, $d$, was added in order to improve the quality of fit.  The coefficients are expanded as follows:

\begin{align}
X=\sum^{4}_{i=0}X_{i}(p_{p}-p_{o})^{i}\mbox{, for }X = a,b,c,d
\end{align}
where $X_{i}$ are the parameters of the fit.  In order to obtain a numerically stable fit, the parameter $p_{0}$ was set to the middle of the momentum range at the center of the carbon block, or 0.55 GeV/c.  The quality of the obtained fits was largely insensitive to the choice of $p_{0}$, provided it was roughly in the middle of the momentum range being fit.  The energy loss was approximated by dividing the carbon block into 1 cm layers and calculating the energy loss at each layer using a parameterization of the proton stopping power in carbon~\cite{pdg}.  The procedure was repeated up to the center of the carbon block resulting in an average total energy loss.  The entire data set found in Tables~\ref{tab:Ay_362} and~\ref{tab:Ay_687} was fit simultaneously and the parameters of this fit can be found in Table~\ref{tab:fit_param}.

To determine the statistical uncertainty of the fit the data points were shifted randomly within their statistical error bars and the resulting distribution was refit.  The procedure was repeated 100 times and the maximum shift from central analyzing power value at each value of $T_{p}$ and $\theta_{fpp}$ was taken as the statistical uncertainty of the fit at that point.  The systematic uncertainty of the fit was determined by fitting the distributions shifted both upward and downward by the systematic uncertainty of the data and the half width at each value of $T_{p}$ and $\theta_{fpp}$ was taken as the systematic uncertainty of the fit at that point.  The statistical and systematic contributions to the total fit uncertainty at each point were then added in quadrature.  The fit uncertainty was then parameterized as a function of $T_{p}$ and $\theta_{fpp}$ using the functional form of equation 5 and the parameters of the error fit can be found in Table~\ref{tab:err_param}.  The uncertainty of the parameterized analyzing power can therefore be calculated for any kinetic energy at carbon center between 82 and 217 MeV and scattering angle between 5 and 41$^{\circ}$.

\section{Results}

\begin{figure*}[f]
    \centering
    \includegraphics[width=14cm]{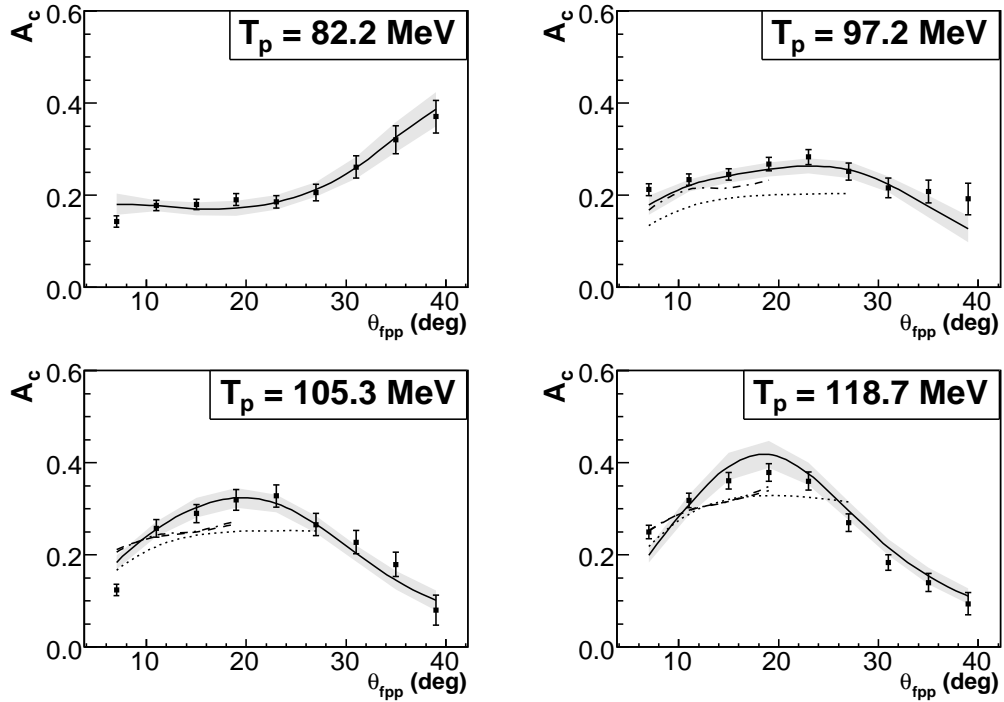}
    \caption[width=12cm]{Analyzing power data and parameterization for beam energy $E_{e}$ = 362 MeV. Error bars shown are statistical and systematic uncertainties added in quadrature.  The dashed line denotes the ``low energy'' McNaughton parameterization~\cite{McNaughton1985}, the dashed dotted the ``low energy'' Aprile-Giboni parameterization~\cite{Aprile}, the dotted line the Waters parameterization~\cite{Waters}, the solid line the new parameterization from this work and the gray area the error band.}
    \label{fig:Ay_362}
  \end{figure*}

\begin{figure*}[f]
   \centering
    \includegraphics[width=14cm]{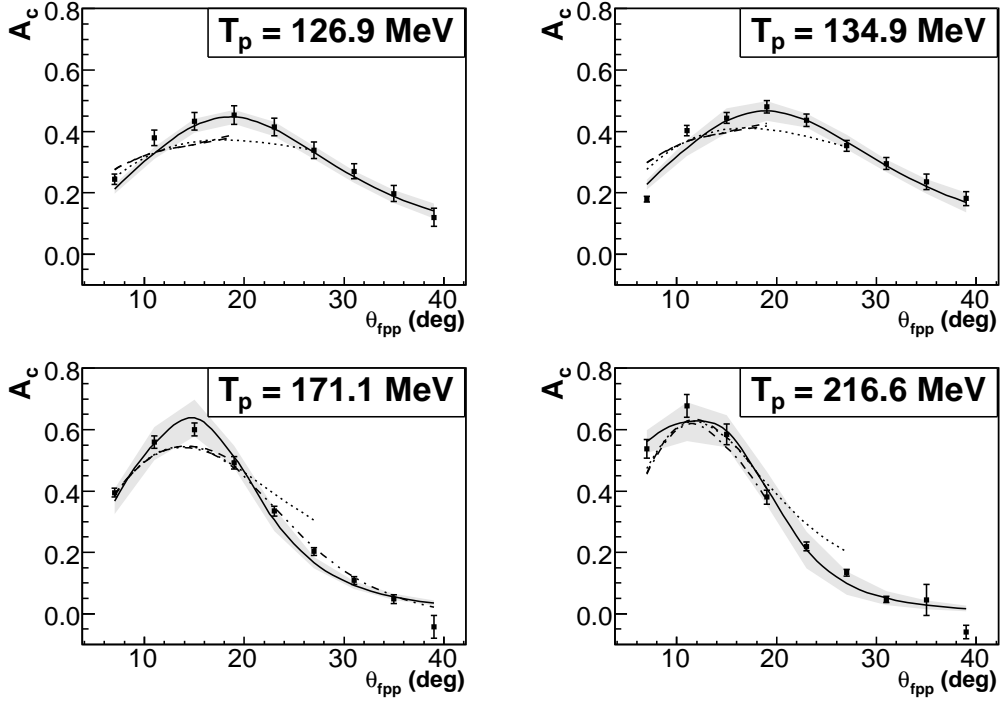}
    \caption[width=12cm]{Analyzing power for beam energy $E_{e}$ = 687 MeV. Error bars shown are statistical and systematic uncertainties added in quadrature.  The dashed line denotes the ``low energy'' McNaughton parameterization~\cite{McNaughton1985}, the dashed dotted the ``low energy'' Aprile-Giboni parameterization~\cite{Aprile}, the dotted line the Waters parameterization~\cite{Waters}, the dashed double dotted line the Pospischil parameterization~\cite{Pospischil}, the solid line the new parameterization from this work and the gray area the error band.}
    \label{fig:Ay_687}
  \end{figure*}

Figures~\ref{fig:Ay_362} and~\ref{fig:Ay_687} compare the new (solid), the ``low energy'' McNaughton~\cite{McNaughton1985} (dashed), the ``low energy'' Aprile-Giboni~\cite{Aprile} (dashed dotted), the Waters~\cite{Waters} (dotted) and the Pospischil~\cite{Pospischil} (dashed double dotted) parameterizations to the new low energy data.  Note that low energy data was also taken by M.~Ieiri {\it et al.}~\cite{Ieiri} for kinetic energies at carbon center between 20 and 84 MeV and scattering angles between 15 and 80$^{\circ}$ and by S.M.~Bowyer {\it et al.}~\cite{Bowyer} for kinetic energies at carbon center between 120 and 200 MeV and scattering angles between 6 and 23$^{\circ}$.  The fit uncertainty of the new parameterization, as described above, is denoted by the gray band.  The ``low energy'' Aprile-Giboni parameterization was done for scattering angles of 5--20$^{\circ}$, energies at carbon center of 90--386 MeV and a carbon block thickness of 3 cm.  The Waters parameterization was done for scattering angles of 3.5--28$^{\circ}$, energies at carbon center of 90--450 MeV and carbon block thicknesses of 3 and 6 cm.  Note that Waters {\it et al.} quote a 10--20\% uncertainty on their fit for energies between 90 and 100 MeV.  The ``low energy'' McNaughton parameterization was fit to the world database at that time, including data from Aprile-Giboni {\it et al.}~\cite{Aprile} and Waters {\it et al.}~\cite{Waters}, with carbon thicknesses ranging from 3 to 12.7 cm.  It is valid for scattering angles between 5 and 20$^{\circ}$ and kinetic energies at the center of the carbon between 100 and 450 MeV. The parameterization of Pospischil {\it et al.}, with a carbon block thickness of 7 cm, used a larger angular range of 15.7--43.1$^{\circ}$ but a narrow energy range of 160--200 MeV.  It is shown only in the $T_{p}$ = 171.1 MeV panel of Figure~\ref{fig:Ay_687}.  

There is a good agreement between the new parameterization and the older ones in the energy/angle regimes for which they were intended, considering all fits were done for different polarimeters and for varying carbon block thicknesses.  Consistency between different polarimeters has been observed before and is generally expected.  This is in part because the energy range covered by most polarimeters in a single measurement is sufficiently small that the variation of analyzing power with energy is essentially linear.  The new parameterization provides an extension down to an energy of 82 MeV and up to a scattering angle of 41$^{\circ}$.  

\begin{figure}[f]
    \centering
    \includegraphics[width=8.8cm]{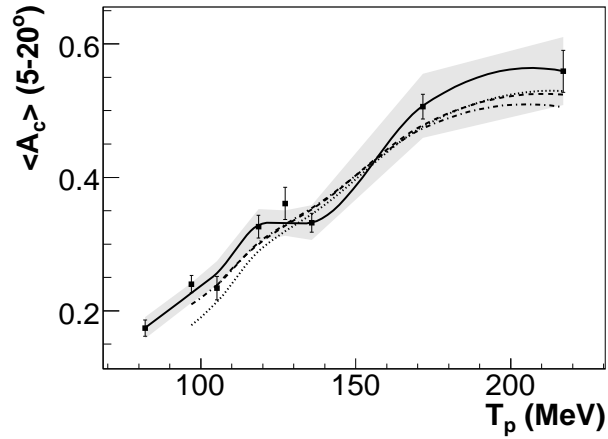}
    \caption[width=12cm]{Weighted average analyzing power as a function of proton energy at the center of the carbon ($T_{p}$) for $5^{\circ} < \theta_{fpp} < 20^{\circ}$. Error bars shown are statistical and systematic uncertainties added in quadrature.  The dashed line denotes the ``low energy'' McNaughton parameterization~\cite{McNaughton1985}, the dashed dotted the ``low energy'' Aprile-Giboni parameterization~\cite{Aprile}, the dotted line the Waters parameterization~\cite{Waters}, the solid line the new parameterization from this work and the gray area the error band.  Note that oscillations in the data near $T_{p}$ = 130 MeV are due to changes in the analyzer thickness.}
    \label{fig:Ay_E_thfpp20}
  \end{figure}

The energy dependence of the weighted average analyzing power is shown for the angular range of $5^{\circ} < \theta_{fpp} < 20^{\circ}$ in Figure~\ref{fig:Ay_E_thfpp20}, which again shows a very good agreement between the new data and the parameterizations in the energy/angle regimes for which the previous parameterizations were intended.  

\begin{figure}[f]
    \centering
    \includegraphics[width=8.8cm]{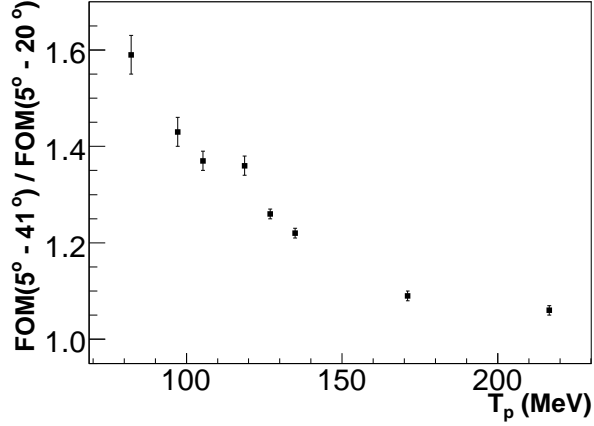}
    \caption[width=12cm]{Figure of merit as a function of maximum scattering angle for beam energy $E_{o}$ = 362 MeV.  Solid line denotes proton kinetic energy at center of carbon analyzer $T_{p}$ = 82.2 MeV, dashed line $T_{p}$ = 97.2 MeV, dotted line $T_{p}$ = 105.3 MeV and dashed dotted line $T_{p}$ = 118.7 MeV.  Note that the figure of merit for $T_{p}$ = 105.3 MeV (0.75 and 2.25'' carbon blocks) is larger than for $T_{p}$ = 118.7 MeV (0.75'' carbon block) because at these energies the figure of merit is greater for a 2.25'' analyzer thickness than a 0.75'' one.}
    \label{fig:fom_thmax_362}
  \end{figure}

\begin{figure}[f]
    \centering
    \includegraphics[width=8.8cm]{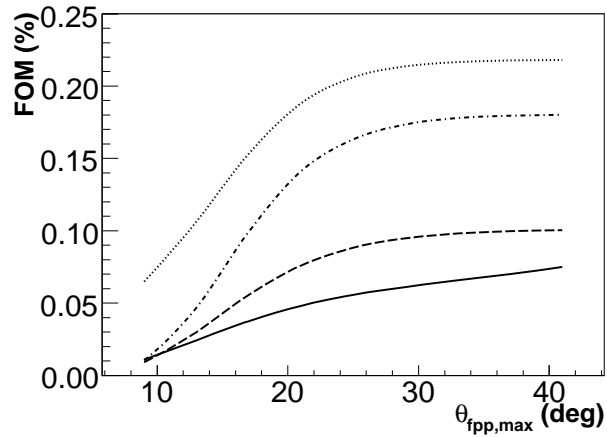}
    \caption[width=12cm]{Figure of merit as a function of maximum scattering angle for beam energy $E_{o}$ = 687 MeV.  Solid line denotes proton kinetic energy at center of carbon analyzer $T_{p}$ = 126.9 MeV, dashed line $T_{p}$ = 134.9 MeV, dotted line $T_{p}$ = 171.1 MeV and dashed dotted line $T_{p}$ = 216.6 MeV.}
    \label{fig:fom_thmax_687}
  \end{figure}

\begin{figure}[f]
    \centering
    \includegraphics[width=8.8cm]{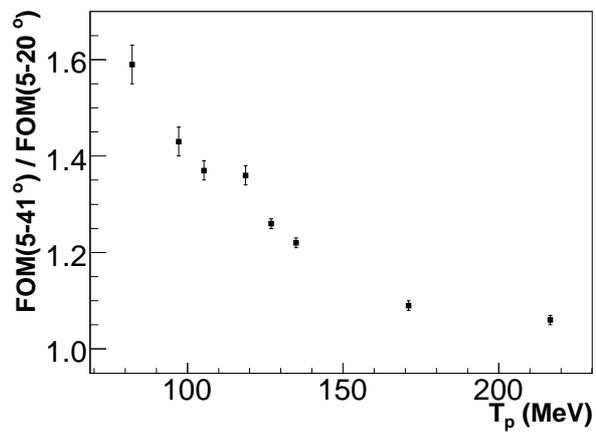}
    \caption[width=12cm]{Figure of merit ratio for two $\theta_{fpp}$ ranges ($FOM(5-41^{\circ})/FOM(5-20^{\circ}$)) as a function of proton energy at carbon center ($T_{p}$).  Uncertainties were calculated using the analyzing power error parameterization and the coefficients in Table~\ref{tab:err_param}.}
    \label{fig:fom_ratio}
  \end{figure}

A higher figure of merit results in more useful events and a lower statistical uncertainty on extracted polarization observables.  The figure of merit as a function of maximum scattering angle is shown in Figures~\ref{fig:fom_thmax_362} and~\ref{fig:fom_thmax_687} for beam energies 362 and 687 MeV, respectively.  For the highest energies, the dependence of the figure of merit on maximum scattering angle is relatively flat beyond 25$^{\circ}$.  For the lowest energies, the figure of merit increases steadily as the maximum scattering angle is increased to 41$^{\circ}$.  By extending the maximum scattering angle of the parameterization, one obtains an improvement of up to 1.6 times in the figure of merit at low energies, as can be seen in Figure~\ref{fig:fom_ratio}.

\section{Conclusion}

A new parameterization, similar to that of~\cite{McNaughton1985}, was determined for p-C analyzing power in the 82 to 217 MeV energy region, and found to be in good agreement with the previous parameterizations in the energy and angular regimes for which they were intended.  This corroborates that the various polarimeters built to date are measuring a property of inclusive proton-carbon scattering insensitive to details of how the polarimeters are constructed.  The extension made for energies between 82 and 90 MeV and for scattering angles between 20 and 41$^{\circ}$ increased the figure of merit by a factor of 1.6 at low energies.  Any new polarimeters for low energy protons should take advantage of the increase in figure of merit from large-angle scattering in the polarimeter.

\section*{Acknowledgments}
We thank the Jefferson Lab physics and accelerator divisions for their contributions.  This work was supported by the U.S. Department of Energy, the U.S. National Science Foundation, Argonne National Laboratory under contract DE-AC02-06CH11357,  the Israel Science Foundation, the Korea Science Foundation, the US-Israeli Bi-National Scientific Foundation, the Natural Sciences and Engineering Research Council of Canada, the Killam Trusts Fund and the Walter C. Sumner Foundation.  Jefferson Science Associates operates the Thomas Jefferson National Accelerator Facility under DOE contract DE-AC05-06OR23177.  The polarimeter was funded by the U.S. National Science Foundation, grants PHY 9213864 and PHY 9213869.

\begin{table}[f]
\begin{tabular}{lccccc}
\hline
$E_{e}$ & $\theta_{p}^{lab}$ & $p_{p}$ & $T_{p}$ & Analyzer \\
  (MeV)    &    (deg)       &  (MeV/c) &  (MeV)  & Thickness \\
           &                 &         &         & (inches) \\
 \hline  
  362 & 28.3 & 476.1 & 82.2 & 0.75 \\
  362 & 23.9 & 503.2 & 97.2 & 0.75 \\
  362 & 18.8 & 526.3 & 105.3 & 0.75  \\
  362 & 18.8 & 526.3 & 105.3 & 2.25 \\
  362 & 14.0 & 541.4 & 118.7 & 0.75\\
  687 & 47.0 & 583.9 & 126.9 & 2.25 \\
  687 & 44.2 & 621.1 & 134.9 & 3.75 \\
  687 & 40.0 & 674.0 & 171.1 & 3.75 \\
  687 & 34.4 & 740.3 & 216.6 & 3.75 \\
\hline
\end{tabular}
\caption{Kinematics of the primary $^{1}$H$(\vec{e},\vec{p})$ reaction and FPP parameters.  $E_{e}$, $\theta_{lab}$ and $p_{p}$ are the beam energy, proton lab angle and proton central spectrometer momentum, respectively.  $T_{p}$ is the proton kinetic energy at the center of the carbon analyzer.  Note that two carbon thicknesses (0.75 and 2.25'') were used for the $\theta_{p}^{lab}$ = 18.8$^{\circ}$ kinematic setting.}
\label{tab:kinematics}
\end{table}

\begin{table}[f]
		\begin{tabular}{lccc}
			\hline
			$T_{p}$ & $\theta_{fpp}$ & $A_{c,i} \pm$ stat $\pm$ syst & $\epsilon_{fpp,i}$ \\
			  (MeV) & (deg) & & (\%)\\
			  \hline
				82.2 & 7	& 0.143	$\pm$ 0.010	$\pm$ 0.007	& 0.33 \\
             & 11	& 0.177	$\pm$ 0.008	$\pm$ 0.008	& 0.42 \\
						 & 15	& 0.180	$\pm$ 0.008	$\pm$ 0.008	& 0.45 \\
						 & 19	& 0.190	$\pm$ 0.009	$\pm$ 0.009	& 0.35 \\
						 & 23	& 0.186	$\pm$ 0.011	$\pm$ 0.008	& 0.22 \\
						 & 27	& 0.206	$\pm$ 0.015	$\pm$ 0.010	& 0.12 \\
             & 31	& 0.261	$\pm$ 0.020	$\pm$ 0.013	& 0.07 \\
             & 35	& 0.320	$\pm$ 0.026	$\pm$ 0.016	& 0.04 \\
             & 39	& 0.371	$\pm$ 0.031	$\pm$ 0.017	& 0.03 \\
			 97.2  & 7	& 0.212	$\pm$ 0.010	$\pm$ 0.009	& 0.28 \\
             & 11	& 0.234	$\pm$ 0.007	$\pm$ 0.010	& 0.41 \\
             & 15	& 0.245	$\pm$ 0.007	$\pm$ 0.010	& 0.41 \\
             & 19	& 0.268	$\pm$ 0.009	$\pm$ 0.012	& 0.29 \\
             & 23	& 0.283	$\pm$ 0.011	$\pm$ 0.012	& 0.17 \\
             & 27	& 0.252	$\pm$ 0.015	$\pm$ 0.011	& 0.10 \\
             & 31	& 0.216	$\pm$ 0.019	$\pm$ 0.011	& 0.06 \\
             & 35	& 0.208	$\pm$ 0.023	$\pm$ 0.009	& 0.04 \\
             & 39	& 0.192	$\pm$ 0.033	$\pm$ 0.010	& 0.03 \\
			 105.3 & 7	& 0.124	$\pm$ 0.008	$\pm$ 0.009	& 1.93 \\
             & 11	& 0.258	$\pm$ 0.009	$\pm$ 0.017	& 0.68 \\
             & 15	& 0.290	$\pm$ 0.009	$\pm$ 0.018	& 0.58 \\
             & 19	& 0.319	$\pm$ 0.011	$\pm$ 0.020	& 0.37 \\
             & 23	& 0.328	$\pm$ 0.014	$\pm$ 0.020	& 0.20 \\
             & 27	& 0.266	$\pm$ 0.017	$\pm$ 0.017	& 0.11 \\
             & 31	& 0.228	$\pm$ 0.021	$\pm$ 0.014	& 0.06 \\
             & 35	& 0.179	$\pm$ 0.024	$\pm$ 0.010	& 0.04 \\
             & 39	& 0.080	$\pm$ 0.032	$\pm$ 0.007	& 0.03 \\
			 118.7 & 7	& 0.249	$\pm$ 0.008	$\pm$ 0.012	& 0.26 \\
             & 11	& 0.318	$\pm$ 0.007	$\pm$ 0.014	& 0.39 \\
             & 15	& 0.361	$\pm$ 0.007	$\pm$ 0.016	& 0.36 \\
             & 19	& 0.379	$\pm$ 0.009	$\pm$ 0.017	& 0.25 \\
             & 23	& 0.360	$\pm$ 0.011	$\pm$ 0.016	& 0.16 \\
             & 27	& 0.270	$\pm$ 0.013	$\pm$ 0.014	& 0.11 \\
             & 31	& 0.184	$\pm$ 0.015	$\pm$ 0.008	& 0.09 \\ 
             & 35	& 0.140	$\pm$ 0.017	$\pm$ 0.009	& 0.06 \\
             & 39	& 0.094	$\pm$ 0.023	$\pm$ 0.008	& 0.03 \\
				  \hline
				\end{tabular}
				 		\caption{Extracted analyzing power for electron beam energy $E_{e}$ = 362 MeV.}
				 		\label{tab:Ay_362}
    \end{table}

    \begin{table}[f]
		\begin{tabular}{lccc}
			\hline
			$T_{p}$ & $\theta_{fpp}$ & $A_{c,i} \pm$ stat $\pm$ syst & $\epsilon_{fpp,i}$\\
			  (MeV) & (deg) & & (\%)\\
			\hline
							126.9 & 7	  &	0.244	$\pm$ 0.006	$\pm$ 0.016	&	1.34 \\
                    &	11	&	0.379	$\pm$ 0.007	$\pm$ 0.024	&	1.24 \\
                    &	15	&	0.433	$\pm$ 0.008	$\pm$ 0.027	&	0.97 \\
                    &	19	&	0.453	$\pm$ 0.010	$\pm$ 0.029	&	0.59 \\
                    &	23	&	0.415	$\pm$ 0.012	$\pm$ 0.026	&	0.34 \\
                    &	27	&	0.339	$\pm$ 0.016	$\pm$ 0.022	&	0.21 \\
                    &	31	&	0.270	$\pm$ 0.018	$\pm$ 0.017	&	0.15 \\
                    &	35	&	0.198	$\pm$ 0.023	$\pm$ 0.013	&	0.10 \\
                    &	39	&	0.120	$\pm$ 0.029	$\pm$ 0.008	&	0.07 \\
					 	134.9   &	7	  &	0.179	$\pm$ 0.004	$\pm$ 0.008	&	2.35 \\
                    &	11	&	0.403	$\pm$ 0.005	$\pm$ 0.016	&	1.47 \\
                    &	15	&	0.443	$\pm$ 0.005	$\pm$ 0.017	&	1.14 \\
                    &	19	&	0.480	$\pm$ 0.007	$\pm$ 0.019	&	0.68 \\
                    &	23	&	0.436	$\pm$ 0.009	$\pm$ 0.018	&	0.38 \\
                    &	27	&	0.353	$\pm$ 0.011	$\pm$ 0.014	&	0.21 \\
                    &	31	&	0.295	$\pm$ 0.016	$\pm$ 0.011	&	0.13 \\
                    &	35	&	0.236	$\pm$ 0.024	$\pm$ 0.009	&	0.08 \\
                    &	39	&	0.181	$\pm$ 0.022	$\pm$ 0.007	&	0.05 \\
					 	171.1   &	7	  &	0.394	  $\pm$ 0.006	$\pm$ 0.013	&	1.38 \\
                    &	11	&	0.559	$\pm$ 0.006	$\pm$ 0.019	&	1.28 \\
                    &	15	&	0.600	$\pm$ 0.007	$\pm$ 0.020	&	0.93 \\
                    &	19	&	0.493	$\pm$ 0.009	$\pm$ 0.018	&	0.60 \\
                    &	23	&	0.334	$\pm$ 0.010	$\pm$ 0.013	&	0.44 \\
                    &	27	&	0.203	$\pm$ 0.011	$\pm$ 0.007	&	0.33 \\
                    &	31	&	0.109	$\pm$ 0.012	$\pm$ 0.004	&	0.25 \\
                    &	35	&	0.048	$\pm$ 0.014	$\pm$ 0.002	&	0.18 \\
                    &	39	&	-0.043	$\pm$ 0.036	$\pm$ 0.004	&	0.11 \\
				    216.6   &	7	  &	0.537	$\pm$ 0.006	$\pm$ 0.030	&	1.23 \\
                    &	11	&	0.677	$\pm$ 0.006	$\pm$ 0.037	&	1.15 \\
                    &	15	&	0.584	$\pm$ 0.007	$\pm$ 0.033	&	0.85 \\
                    &	19	&	0.380	$\pm$ 0.008	$\pm$ 0.022	&	0.67 \\
                    &	23	&	0.219	$\pm$ 0.008	$\pm$ 0.012	&	0.59 \\
                    &	27	&	0.134	$\pm$ 0.008	$\pm$ 0.008	&	0.53 \\
                    &	31	&	0.047	$\pm$ 0.009	$\pm$ 0.003	&	0.41 \\
                    &	35	&	0.045	$\pm$ 0.048	$\pm$ 0.017	&	0.30 \\
                    &	39	&	-0.060	$\pm$ 0.021	$\pm$ 0.003	&	0.18 \\
					 \hline
		\end{tabular}
		 					 \caption{Extracted analyzing power for electron beam energy $E_{e}$ = 687 MeV.}
		 					 \label{tab:Ay_687}
\end{table}

\begin{table}[f]
\begin{tabular}{lcccc}
\hline
 $T_{p}$  & $\epsilon_{fpp}$ & $<A_{c}> \pm stat \pm syst$ & $FOM$ \\
  (MeV)   & (\%)             &                             & (\%)\\
  \hline  
82.2  & 2.03 &  0.186 $\pm$ 0.010 $\pm$ 0.009 & 0.07 \\
97.2  & 1.80 &  0.243 $\pm$ 0.010 $\pm$ 0.010 & 0.11\\
105.3 & 4.01 &  0.239 $\pm$ 0.011 $\pm$ 0.015 & 0.20\\
118.7 & 1.72 &  0.307 $\pm$ 0.009 $\pm$ 0.014 & 0.17\\
126.9 & 5.02 &  0.354 $\pm$ 0.009 $\pm$ 0.023 & 0.66 \\
134.9 & 6.50 &  0.336 $\pm$ 0.006 $\pm$ 0.014 & 0.81\\
171.1 & 5.49 &  0.423 $\pm$ 0.009 $\pm$ 0.015 & 1.16\\ 
216.6 & 5.91 &  0.391 $\pm$ 0.010 $\pm$ 0.023 & 1.31\\
\hline
\end{tabular}
\caption{Weighted average analyzing power ($<A_{c}>$), FPP efficiency ($\epsilon_{fpp}$) and figure of merit ($FOM$).  The $\epsilon_{fpp}$ values are sums of $\epsilon_{fpp}^{i}$ over all bins in $\theta_{fpp}$ in Tables~\ref{tab:Ay_362} and~\ref{tab:Ay_687}.  The $FOM$ values were calculated using the summation approximation of equation~\ref{eq:fom} and, along with $<A_{c}>$, the entire angular range of $5^{\circ} < \theta_{fpp} < 41^{\circ}$.}
\label{tab:Ay_average}
\end{table}

\begin{table}[f]
\begin{tabular}{lll}
\hline
Coefficient & Value & Unit\\
            &       & \\
\hline
p$_{0}$ & 0.55 & (GeV/c) \\
a$_{0}$ & 4.0441 & (GeV/c)$^{-1}$\\
a$_{1}$ & 19.313 & (GeV/c)$^{-2}$\\
a$_{2}$ & 119.27 & (GeV/c)$^{-3}$\\
a$_{3}$ & 439.75  & (GeV/c)$^{-4}$\\
a$_{4}$ & 9644.7 & (GeV/c)$^{-5}$\\
b$_{0}$ & 6.4212 & (GeV/c)$^{-2}$\\
b$_{1}$ & 111.99 & (GeV/c)$^{-3}$\\
b$_{2}$ & -5847.9  & (GeV/c)$^{-4}$\\
b$_{3}$ & -21750  & (GeV/c)$^{-5}$\\
b$_{4}$ & 973130 & (GeV/c)$^{-6}$\\
c$_{0}$ &  42.741 & (GeV/c)$^{-4}$\\
c$_{1}$ & -8639.4& (GeV/c)$^{-5}$\\
c$_{2}$ & 87129 & (GeV/c)$^{-6}$\\
c$_{3}$ & 8.1359 $ \times 10^{5}$ & (GeV/c)$^{-7}$\\
c$_{4}$ & -2.1720 $\times 10^{7}$ & (GeV/c)$^{-8}$\\
d$_{0}$ &  5826.0 & (GeV/c)$^{-6}$\\
d$_{1}$ &  2.4701$\times 10^{5}$   & (GeV/c)$^{-7}$\\
d$_{2}$ & 3.3768 $ \times 10^{6}$ & (GeV/c)$^{-8}$\\
d$_{3}$ & -1.1201$ \times 10^{7}$ & (GeV/c)$^{-9}$\\
d$_{4}$ & -1.9356$ \times 10^{7}$ & (GeV/c)$^{-10}$\\

\hline
\end{tabular}
\caption{Parameters of analyzing power polynomial fit; reduced $\chi^{2}$ of the fit is 1.39 with a $\chi^{2}$ of 71.0 and 51 degrees of freedom.}
\label{tab:fit_param}
\end{table}

\begin{table}[f]
\begin{tabular}{lll}
\hline
Error       & Value & Unit \\
Coefficient &       &   \\
            &       & \\
\hline
p$_{0}^{e}$ & 0.55 & (GeV/c) \\
a$_{0}^{e}$ & 0.34687 & (GeV/c)$^{-1}$\\
a$_{1}^{e}$ & 3.8266 & (GeV/c)$^{-2}$\\
a$_{2}^{e}$ & 11.635 & (GeV/c)$^{-3}$\\
a$_{3}^{e}$ & -314.18  & (GeV/c)$^{-4}$\\
a$_{4}^{e}$ & 1015.1 & (GeV/c)$^{-5}$\\
b$_{0}^{e}$ & 0.21579 & (GeV/c)$^{-2}$\\
b$_{1}^{e}$ & 288.29 & (GeV/c)$^{-3}$\\
b$_{2}^{e}$ & -13736.  & (GeV/c)$^{-4}$\\
b$_{3}^{e}$ & -1.5821 $\times 10^{5}$  & (GeV/c)$^{-5}$\\
b$_{4}^{e}$ & 2.2523  $\times 10^{6}$ & (GeV/c)$^{-6}$\\
c$_{0}^{e}$ &  938.40 & (GeV/c)$^{-4}$\\
c$_{1}^{e}$ & -4961.8 & (GeV/c)$^{-5}$\\
c$_{2}^{e}$ & 4.0190 $\times 10^{5}$& (GeV/c)$^{-6}$\\
c$_{3}^{e}$ & 3.3034 $\times 10^{6}$ & (GeV/c)$^{-7}$\\
c$_{4}^{e}$ & -5.8401$\times 10^{7}$ & (GeV/c)$^{-8}$\\
d$_{0}^{e}$ &  -4910.9 & (GeV/c)$^{-6}$\\
d$_{1}^{e}$ &  1.5133 $\times 10^{5}$ & (GeV/c)$^{-7}$\\
d$_{2}^{e}$ & -1.3127 $ \times 10^{6}$ & (GeV/c)$^{-8}$\\
d$_{3}^{e}$ & -2.7446$ \times 10^{7}$ & (GeV/c)$^{-9}$\\
d$_{4}^{e}$ & 2.8556$ \times 10^{8}$ & (GeV/c)$^{-10}$\\
\hline
\end{tabular}
\begin{flushleft}
\caption{Parameters of fit to the uncertainty in analyzing power parameterization.}
\end{flushleft}
\label{tab:err_param}
\end{table}

\end{document}